# Harnessing Poly(ionic liquid)s for Sensing Applications

Ryan Guterman, Martina Ambrogi, Jiayin Yuan[*]

Dr. R. Guterman, Dr. M. Ambrogi, Dr. J. Yuan

Department of Colloid Chemistry, Max Planck Institute of Colloids and Interfaces

Am Mühlenberg 1 OT Golm, D-14476 Potsdam, Germany

E-Mail: jiayin.yuan@mpikg.mpg.de

**Abstract:**

The interest in poly(ionic liquids) for sensing applications are derived from their strong interactions to a variety of analytes. By combining the desirable mechanical properties of polymers with the physical and chemical properties of ILs, new materials can be created. The tunable nature of both ionic liquids and polymers allows for incredible diversity, which is exemplified in their broad applicability. In this article we examine the new field of poly(ionic liquid) sensors by providing a detailed look at the current state-of-the-art sensing devices for solvents, gasses, biomolecules, pH, and anions.

TOC image

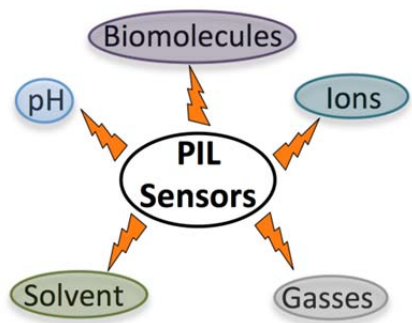

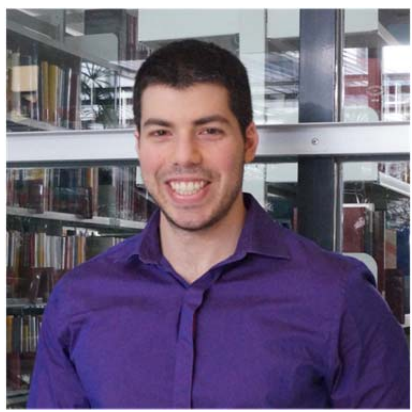

Dr. Ryan Guterman acquired his BSc and PhD in chemistry at The University of Western Ontario (UWO), Canada in 2010 and 2015 respectively. In October of 2015 he joined the Max Planck Institute of Colloids and Interfaces in Potsdam as a Postdoctoral Fellow, working with Dr. Jiayin Yuan and Dr. Markus Antonietti. His current interests includes inorganic chemistry, material science, and the design of new poly(ionic liquid)s.

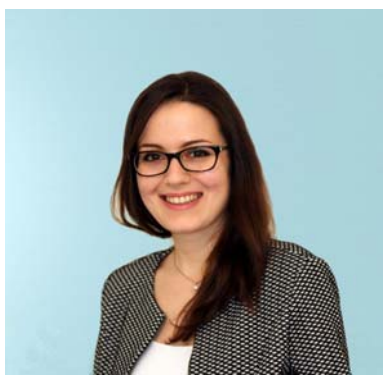

Dr. Martina Ambrogi completed her studies at the university of Bologna, Italy, where she graduated in 2012 as MSc in industrial chemistry. In 2013 she joined Dr. Jiayin Yuan group at the Max Planck Institute of Colloids and Interfaces in Potsdam, under the supervision of Prof. Dr. Markus Antonietti. Being part of a European Union – ITN Marie Curie Action project as early stage research, she focused on the use of poly(ionic liquid)s for energy applications, successfully achieving her PhD in polymers and colloids chemistry in 2015.

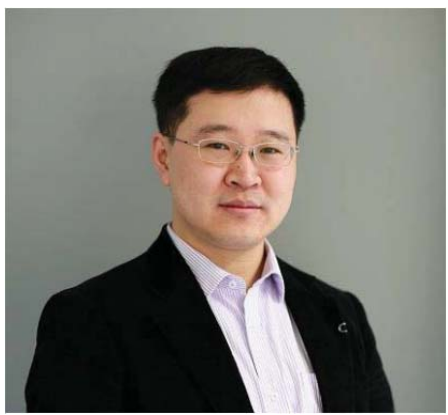

Dr. Jiayin Yuan studied chemistry in Shanghai Jiao Tong University, China in 1998. He received his master degree in University of Siegen in 2004, and doctoral degree (summa cum laude) in University of Bayreuth, Germany in 2009. He then joined the Max Planck Institute of Colloids and Interfaces in Potsdam, where he was promoted to a research group leader in 2011. He received the ERC Starting Grant in 2014 and his current interest focuses on poly(ionic liquid)-derived functional (nano)materials.

The concept of designing functional polymers from ionic liquids (ILs) has recently attracted rapidly increasing interest in the field of polymer and materials science. The amalgamation of IL chemistry, physics, and polymers manifests in a variety of new research topics and discoveries. A frontier topic in this regard is poly(ionic liquid)s, or polymerized ionic liquids (PILs), which in the strict sense are the polymerization products of ILs. Polymers synthesized *via* chemical modification of existent polymers through ion exchange, *N*-alkylation, or ligation of ILs onto polymers have also been classified as PILs by some groups.[1,2] Therefore in a broad sense, this class of materials refers to functional ionic polymers that carry a high density of covalently assembled IL through a polymeric backbone or framework. The unique structural configuration gives rise to a group of polymeric materials that currently broadens the property and applicative spectra of ILs and conventional polyelectrolytes. Such an innovation in polymer design has created new fundamental and practical research frontiers, including antibacterial coatings,[3,4] ion conductive films,[5–7] porous polyelectrolyte membranes,[8–10] "smart" dispersants,[11,12] and heteroatom-doped carbon nanomaterials.[13–15] A particular characteristic of PIL research is its interdisciplinary nature, where scientists of various backgrounds join together to promote this topic in different dimensions. While there are a number of elegant and pioneering PIL-related review papers over the past 5 years that examine the diverse perspectives of PILs,[16–24] our primary goal in this contribution is to highlight the emerging trend in investigating the sensing potential of PILs.

The development of high-quality sensors that are cost-effective often requires novel materials and design principles. One such material that has received attention are ILs, traditionally defined as molten salts with melting point below 100°C. ILs have been explored for numerous applications over the past 20 years for catalysis,[25] sensing,[26,27] and $CO_2$ capture.[28] Despite this, the intrinsic poor mechanical properties ILs possess is a visible bottleneck restricting their applicability in fabricating sensors of well-defined shapes, homogeneous coatings, patterned surfaces, fibers or films, where polymeric materials may be more desirable. A solution to this problem can be achieved *via* the use of PILs,[16] which harnesses some of the distinctive physical and chemical properties of ILs with the desirable mechanical properties and processability of polymers. Despite the rapid advance of PILs being used in subfields of materials science, their use in sensing devices has not been considered as a significant branch to be investigated, but rather a by-product of their structures or other functions. Though PILs have long been adopted as structural elements in some sensors, a comprehensive view of their sensing function is currently missing and the "trial and error" principle dominates the design spirit of PIL-containing sensors.

The diversity in the chemical structure of PILs on both the molecular and macromolecular scale is a key to their exploitation in sensing devices. On the molecular scale, various cations and anions may be used which affect the physical and chemical interactions of the polymer with guest molecules. The polymerization of the cation is at present the most common approach for PIL synthesis, and mainly includes imidazolium, pyridinium, ammonium, and phosphonium structures. Both chain-growth[16] and step-growth polymerizations[29,30] have been developed, however free-radical approaches are by far the most common. The toolbox of common anions is much larger than the cations, which ranges from small halides to bulky polyatomic hydrophobic species. The combination of cations and anions provides a multitude of opportunities to precisely tune and engineer the PIL-guest interplay for sensing operation.[31] On the macromolecular scale, the covalent immobilization of the cation-anion pair to a polymer chain or network confines their interaction with an analyte to a well-defined, addressable location. The macroscopic shape of the PIL may, if necessary, be preserved during the sensor operation period because of their polymer-like mechanical properties, which is in stark contrast to ILs. New material designs may be realized through controlled polymerization techniques, including the exploitation of self-assembled systems and nanomaterials with new responsive behaviors.[32,33] For example, PIL star block copolymers displayed reversible temperature-responsive self-assembly, whose behavior was dependent on the location of the PIL block within the macromolecule.[34] Other stimuli that have been explored includes pH,[35,36] light,[37] solvent mixtures,[38,39] and ionic strength,[40] which can induce shape deformations or triggered swelling/deswelling effect.

While much research to date has simply explored the responsive nature of PILs,[41] a phenomenon occurring on a molecular level, their extension into a macroscopically visible/detectable signal is currently under development. The diversity of PIL responsiveness and the extensive toolbox of IL structures suggest they hold great promise for use as sensors. Here we highlight several key applications for PILs in sensing devices, ranging from ion detection, gas sensing, solvent determination, to biosensors, their detection mechanisms, and PIL function.

**Ion sensing**

The advantage of using PILs as an ion sensor lies in their capability to transport, interact or exchange with ions in bulk or in solution. It is well known that the properties of ILs and their polymers are sensitive to subtle changes in anion and cation structure. Since most PILs are polycations, anion detection is a natural instinct of PILs and is among the most explored ion sensors. In the absence of any phase change (e.g. precipitation or phase separation), ion exchange is a spontaneous process occurring on a molecular level when PILs meet ions in their solution. To produce a detectable signal, the change in PIL ion composition should be able to trigger a variation in their physical properties. Given that most PILs are at present imidazolium derived, there has been great interest in understanding imidazolium-anion interactions and the effects they have on small-molecule and PIL solubility.[41] The ion-pair interaction is both sensitive to cation/anion choice but also solvent type, which leads to fertile opportunities for IL-derived anion sensors. For example, Li *et al.* took advantage of the tunable swelling properties PILs possess with different anions to create a simple anion sensor.[42,43] The acrylate-appended imidazolium monomers were copolymerized with a diacrylate crosslinker in the presence of monodispersed colloidal silica spheres, which are 260 nm in size and organized in a densely packed, regular array (Figure 1A). The silica template was subsequently removed by HF treatment to create a porous PIL film with an inverse opal structure. This material acted as a photonic crystal whose color depended on the spacing between the periodic porous PIL network. This approach is one of many to create photonic crystals that can tune the properties of light and is naturally suited for sensing applications, as described by Texter *et al.*.[44] Simply by replacing the original bromide anion by a different type, the periodic spacing in the polymer matrix grew or shrunk due to the swelling or deswelling of the PIL network in water, thus altering the color of the porous material. Hydrophilic anions can swell the PIL network significantly in water, thus expanding the periodic spacing to result in a higher wavelength color shift (Figure 1B and D), while hydrophobic anions result in a smaller spacing and a blue-shift (Figure 1C and D). This process is highly specific to the type of anion used, as each anion dictates a defined swelling degree and color change, leading to excellent anion discrimination. They also discovered that this sensor is not sensitive to pH and can be reused after a sodium hydroxide treatment.

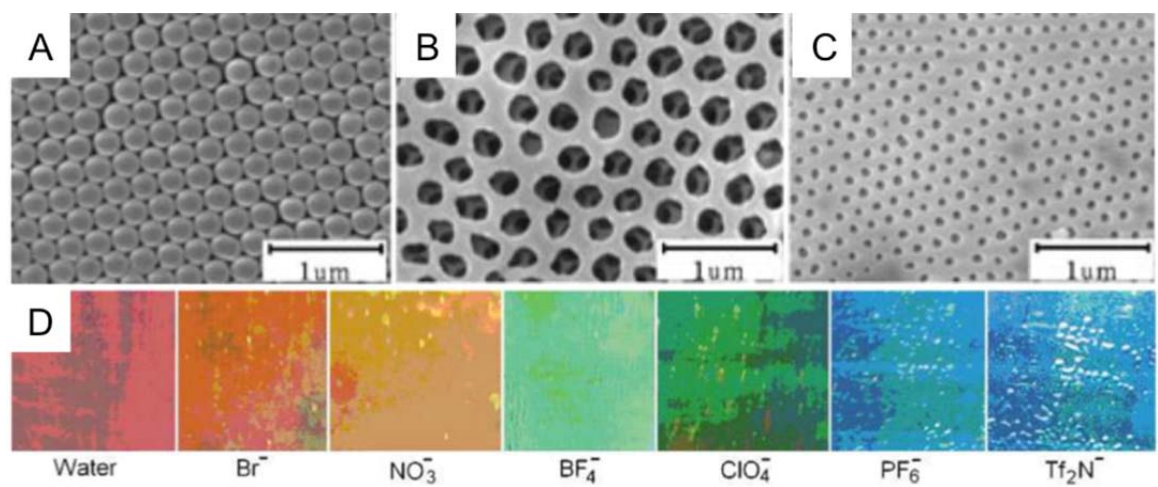

**Figure 1:** A) Regular array of monodispersed silica particles as a photonic crystal. B) Opal structure of Br⁻ containing porous PIL network after HF etching to remove silica template. C) Anion exchange of structured PIL with NTf$_2^-$. D) The induced color changes of the photonic PIL film upon soaking in a variety of aqueous salt solutions (0.02 M).

It is known that swellability of polyelectrolytes does change slightly with pH,[45,46] however these changes might not be dramatic enough to be visualized by the inverse opal, especially in the middle pH range where the corresponding variation in the ionic strength is rather small. Tracing changes in pH requires either more sophisticated exquisite techniques, or a different sensing mechanism beyond the anion-relevant swellability of the PIL network. It has been reported that a fiber-optic interferometer sensor, built up *via* surface coating a thin-core optic fiber with a nanoporous PIL membrane, could precisely measure the pH value in aqueous solution.[47] This was accomplished by monitoring the variation of the optical signal transmitting through the fibers soaked in an aqueous solution. The PIL-based porous membrane skin was constructed *via* interpolyelectrolyte complexation between a cationic PIL poly[3-cyanomethyl-1-vinylimidazolium bis(trifluoromethanesulfonyl)imide] and a neutralized PAA. Since the electrostatic complexation between PIL and PAA cannot reach its full extent, the porous membrane was equipped with residual free carboxylate groups that did not join the electrostatic complexation, which can undergo reversible (de)protonation. In an aqueous environment, the refractive index of the porous membrane changes with pH because of reversible carboxylate ionization, which can be detected as an optical signal. The employment of this nanoporous PIL membrane is crucial for high-performance pH sensing and breaks the so-called "trade-off" rule in sensing function, *i.e.* harnessing high sensitivity and fast response simultaneously. The high sensitivity relies on the thick coating of the membrane onto the fiber surface (50-100 μm in thickness) that amplifies the interaction of the membrane with the solute, while the fast response is a nature outcome of the porous structure to accelerate the diffusion of the solute into the membrane.(Figure 2A).

Colourmetric pH sensing using imidazolium-based PILs has been developed by Yan *et al.* using a pH sensitive anion (cresol red) in a crosslinked gel (Figure 2B and C).[36] Different from the sensing mechanisms of the aforementioned 2 examples, where the reversible physical/chemical interaction of a PIL network with the solutes determines the detection process, the role of PILs here is to serve as a structural scaffold for the pH-responsive anions to fulfill the sensing function. The authors found their approach yielded reusable sensor strips and operated in both aqueous and organic media. A similar method has been investigated by Khan *et al.* using PIL microbeads embedded with a pH sensitive indicator.[48]

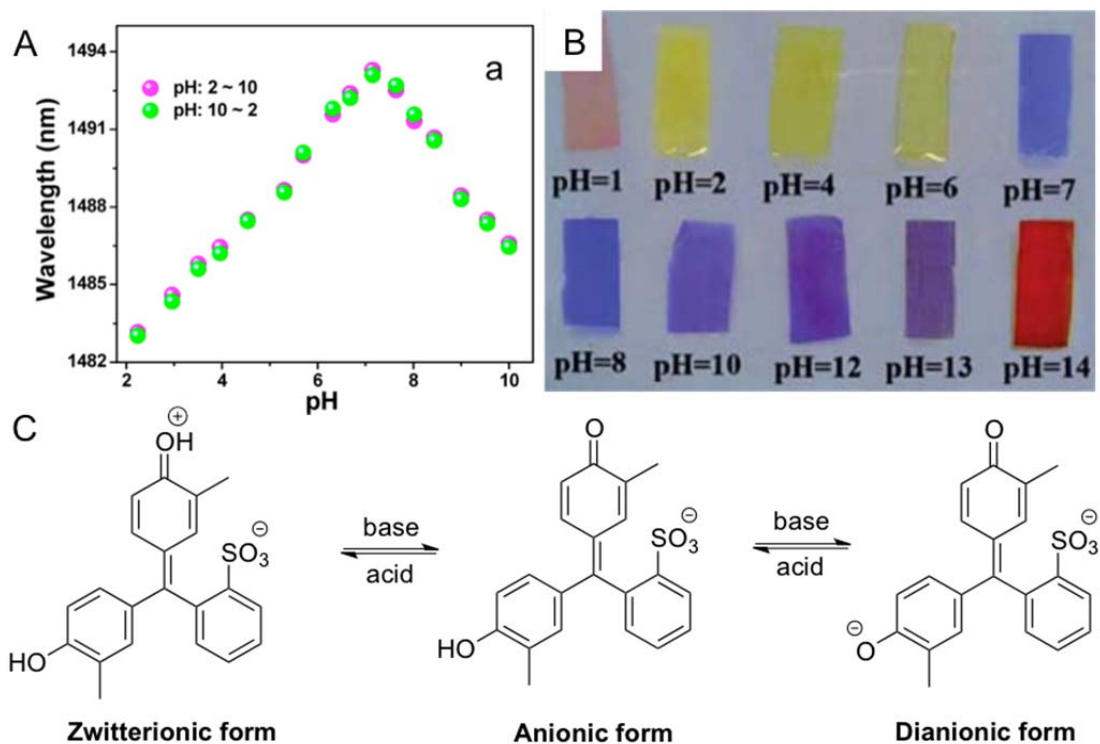

**Figure 2:** A) Reproducible pH detection using a porous PIL-functionalized fiber optic. B) Colourmetric pH sensing using a crosslinked PIL with a cresol red anion. C) Different ionization states of cresol red.

In a more rare example, a self-assembled PIL nanoparticle embedded with a fluorescent dye molecule has been used for the selective detection of $Cu^{2+}$ ions in the presence of $Ca^{2+}$, $Mg^{2+}$, $Zn^{2+}$. The sensing process relies on the quenching effect of the $Cu^{2+}$ ions on the dye molecules.[49] This study is the only example of a PIL sensor for cationic metallic species.

## Gas sensing

The well-known high affinity between polar gasses, such as $CO_2$, and ILs has opened up new applications, ranging from extraction methods,[50,51] $CO_2$ capture,[52,53] and gas separation.[54,55] Depending on the chemical structure of the anion, and to a lesser extent, the cation, large variability in $CO_2$ sorption capacity has been recorded.[28,56] The solubility of $CO_2$ in ILs is derived from the organization of $CO_2$ molecules around the anion, with large fluorinated anions providing the greatest solubilities.[57,58] The sorption of $CO_2$ into ILs also varies some of their physical properties, such as their polarity and conductivity. The interplay between $CO_2$ and ILs sets the base of using PILs for $CO_2$ sensing.

It was found that PILs afford greater $CO_2$ absorption capacities and rates relative to their monomeric ILs,[59,60] with 90% of its capacity achieved in a few minutes for PILs *vs.* several hours for their corresponding IL.[61] These properties have been exploited for $CO_2$ sensing by harnessing the affinity between $CO_2$ and PILs.[62] Jin *et al.* described the synthesis of a PIL/single walled carbon nanotubes (SWNT) composite that was capable of achieving very low detection limits (Figure 3A).[63] The PIL employed here is poly(3-ethyl-1-vinylimidazolium tetrafluoroborate), which attaches to the surface of SWNTs *via* cation-π interaction. The sensing process relies on the electrochemical impedance spectroscopy measurement of the composite upon exposure to a $CO_2/N_2$ mixture, which revealed a decrease in resistance upon increasing $CO_2$ concentration (Figure 3B). The authors pointed out the Lewis acid–base interaction between $BF_4^-$ and $CO_2$ is the sensing mechanism. When $BF_4^-$ anions on SWNT surface encounter $CO_2$, a charge transfer from $BF_4^-$ to $CO_2$ occurs, resulting in a reduction of electron-donation from $BF_4^-$ to the carbon nanotubes, thus increasing the hole population in SWNTs. Even in the presence of other gasses and humidity, the sensor operated with good selectivity. To regenerate the sensor, UV irradiation under a nitrogen atmosphere was required to ease the desorption of $CO_2$ gas molecules from PIL/SWNT surface without affecting the reproducibility of the sensor for further measurements. In spite of high sensitivity and the lowest detection limit (500 ppt) of this sensing system, the setup suffers from a saturation of $CO_2$ concentration up to 50 ppm, a value which is much lower than the ambient $CO_2$ concentration, thus lacks of practical applicability.

To improve the usability of $CO_2$ sensor at ambient environment, a PIL comprised of poly[(*p*-vinylbenzyl)trimethylammonium hexafluorophosphate] in conjunction with $La_2O_2CO_3$ nanoparticles was processed into a thin film that can detect $CO_2$ with good sensitivity in an aerobic, humid atmosphere.[62] To fabricate the sensor, the composite thin film was placed between two electrodes and their resistance was measured by electrochemical impedance spectroscopy (Figure 3C). At an optimum weight ratio of PIL and nanoparticle (60-80 wt% $La_2O_2CO_3$), the interfacial interaction between PIL and the nanoparticles was maximized and resulted in a significant drop in resistance through the formation of charge transfer highways.[64,65] Additionally, the interaction between the negatively charged particles and the cationic PIL resulted in increased anion mobility and thus a further decrease in resistance (Figure 3D). It was found that exposure to $CO_2$ induced a further decrease in the resistance assigned to the conductivity channels at the PIL/nanoparticle interface. This phenomenon is a result of the cationic components in the PIL being sequestered by $CO_2$, resulting in greater mobility of $PF_6^-$. Good sensitivity in a broad $CO_2$ concentration range (150-2400 ppm) that covers the entire concentration range of atmospheric $CO_2$ and high reproducibility was achieved at room temperature with satisfactory response rates (Figure 3E). As for further improvement of the sensing performance, any means to expand the interfacial interaction to reduce the electric resistance as well as to accelerate the $CO_2$ diffusion in the composite can be exerted. For example, by increasing the porosity of the PIL material, greater $CO_2$ adsorption capacities may be obtained thus providing a potential means for greater sensitivity and faster response.[66]

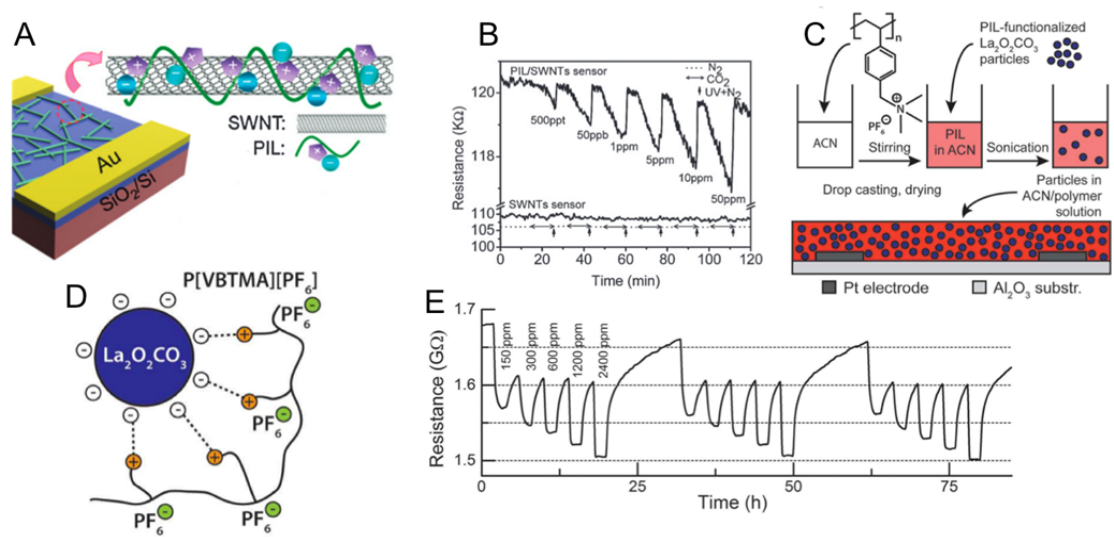

**Figure 3:** A) A $CO_2$ sensor comprised of a PIL-SWNT composite between two gold electrodes. B) Detection of $CO_2$ at ppt concentrations for a PIL-SWNT sensor. C) Formation of a PIL-$La_2O_2CO_3$ composite for $CO_2$ sensing. D) Sequestration of the $La_2O_2CO_3$ allowing for greater $PF_6^-$ mobility. E) Changes in resistance of the PIL-$La_2O_2CO_3$ upon introduction of $CO_2$ gas.

## Solvent sensing

The ability to detect and differentiate solvents in a simple, rapid fashion is a challenge today. While techniques such as nuclear magnetic resonance spectroscopy, mass spectrometry and chromatography can be used to obtain quantitative data, they are generally expensive, time-consuming and impractical for field applications. Accordingly, a large variety of alternative solvent-sensing methodologies have been developed and often rely on simple feedback mechanisms, such as changes in colour, shape and conductivity.[67–69] From this standpoint, PIL-derived soft actuators are naturally ideal candidates, which undergo visible macroscopic locomotion by external stimuli. The intrinsic ionic nature of PILs has made them popular as building blocks to fabricate electroactive actuators, in which the deflected shapes of a film actuator may be a function of voltage applied.[70–73] Nevertheless, the practical devices suffer from dilemmas, such as limited voltage range and back-relaxation, thus their sensing potential is so far restricted. In contrast, soft actuators based on porous PIL membrane do find great potential in sensing appication. Recently a porous PIL membrane with pore size between 0.3 - 3 μm has been developed that displayed excellent bending actuation behaviour upon exposure to vapors of various solvents.[8,74] Physical movement of the membrane is a result of a unique crosslink density gradient through the cross-section of the material that is naturally formed during membrane fabrication (Figure 4A). The asymmetric profile in the crosslinking density results in unbalanced swelling from the membrane top and bottom areas, thus bending the membrane in a predictable direction. The high porosity coupled with the high solvent affinity of the PIL results in ultrafast swelling upon exposure to vapours, and deswelling when unexposed. The sensing mechanism stems from the bending degree or curvature of the membrane controlled by the IL-solvent interaction, thus solvent-specific. From the final state of the curved membrane, one can discriminate between vapours. The initial investigations for solvent sensing demonstrated that polar aprotic solvents (such as acetonitrile, THF, and acetone), which are capable of dissolving the PIL polymer, resulted in larger curvature in the membrane material (Figure 4B). By further replacing the $Tf_2N$ anion with $PF_6$ in the PIL, polar aprotic solvents, such as acetone and acetonitrile vapours can be distinguished quite easily (Figure 4C). The reason for this stems from the different solubilities of the PIL bearing $NTf_2$ *vs* $PF_6$ in acetone. While acetone and acetonitrile can both dissolve the $NTf_2$ PIL, acetone is a poor solvent for the $PF_6$ PIL resulting in less actuation. This demonstrates that tunable and sensitive nature of these PIL membranes for selective solvent sensing. Nevertheless, the sensing mechanism here has to be coupled with the vapor pressure of individual solvents, as high-boiling point solvents, such as DMSO, despite of their strong interactions with the IL species, failed to drive the membrane actuation due to a low concentration in the vapor phase.

When the pore size of the PIL membrane reaches the nanometer scale, the sensing of solvents can be directly performed in their liquid state with extremely high sensitivity, instead of the vapour phase. For example, the nanoporous PIL membrane can sense the concentration of acetone in an aqueous solution with the detection limit at an acetone concentration as low as 0.25 mol%, that is, one acetone molecule per 400 water molecules. Such high sensitivity is one order of magnitude higher than recently reported gel systems.[75] It is also shown that even organic isomers may be distinguished by this method because of their slightly different affinities to the IL species in the PIL membrane. One example reported recently is the 3 isomers of butanol. The least polar isomer, isobutanol, displayed the lowest actuation effects, while 2-butanol and 1-butanol induced greater bending at the same concentration in aqueous phase (Figure 4D).[74]

Feller *et al.* have developed an electrochemical vapor sensor using a silver nanoparticle-reduced graphene oxide/PIL nanocomposite.[76] The PIL assisted in both binding the silver nanoparticles to the graphene, and also dispersing the composites in organic solvent. These stable dispersions were spray-coated onto interdigitated electrodes and the resistance measured upon exposure to a variety of solvent vapours, including methanol, ethanol, methylacetate, acetone, and water. In each case, the resistance increased upon exposure with the strongest signal resulting from methanol. The sensor was only partially sensitive to non-polar compounds, such as chloroform, dichlorobenzene, toluene and styrene. The sensitivity for polar solvents is a result of their strong interaction with the IL species in the PIL component, which disrupts the flow of the electrons through the composite system and increases the resistivity.[77,78] They found that the presence of both the PIL and silver nanoparticles were crucial for optimum device performance. These sensors were reusable, sensitive between 1-200 ppm for methanol vapours, and could easily distinguish between various different solvents.

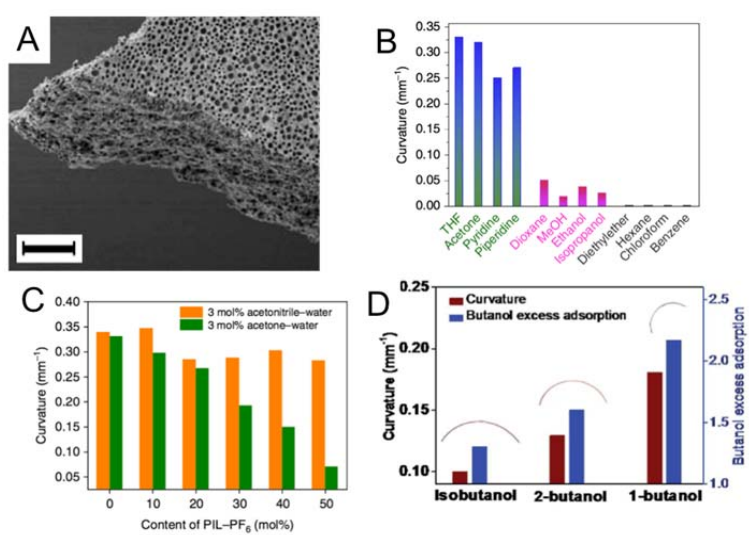

**Figure 4:** A) PIL-polyacid membrane with a gradient porous structure. B) Actuation response to various solvent. Polar aprotic solvents performed the best, followed by polar protic and finally non-polar. C) Actuation response of PIL membranes with varying amount of $PF_6^-$ vs $NTf_2^-$ to either acetone or acetonitrile (3 mol%) in water. D) Sensing different butanol isomers using a porous PIL actuator. The differences in polarity results in different degrees of swelling in the membrane thus resulting in differences in actuation.

Li *et al.* developed a humidity sensor based on the "inverse opal structure" as seen in Figure 1B.[79] Like the anion-detecting system, this material acted as a photonic crystal whose color depended on the periodic spacing between the PIL network.

Simply by changing the relative humidity (RH) from 20% to 100% and back again, the porous network left by the silica spheres in the polymer matrix grow or shrink, thus altering the color of the material. Moisture can swell the PIL resulting in an expansion of the periodic spacing, turning the film from blue at 20% RH, to red at 100% RH. They found that upon switching the anion from bromide to dodecylsulfate, their sensor also responded to ethanol vapours. Another simple electrochemical humidity sensor by Texter *et al.* was created from a crosslinking IL monomer, where both the imidazolium unit and the anion possessed an acrylate functional group.[80] Upon polymerization, the charge-dense material was found to have a very high dielectric constant over a wide frequency range. In the presence of moisture (0-75% RH), the permittivity of the material increased exponentially and was found to be reversible, thus demonstrating a highly sensitive and straightforward approach to humidity sensing.

**Bio-sensor**

Determining the composition and concentration of biological molecules *in vitro* and *in vivo* are crucial for biotechnology and medical applications. One common method is to incorporate PILs into biosensors through the formation of a PIL composite with a conductive entity. For example, the work by Firestone *et al.* and Song *et al.* incorporated PILs with gold nanoparticles[81] and graphene,[82] respectively, to create a glucose sensor. Firestone's approach was to photopolymerize a vinylimidazolium-based IL monomer bearing a long alkyl tail on the cation in the presence of $HAuCl_4$ to create channels of gold nanoparticles which acted as conductive conduits.[83,84] The amphiphilic IL monomer served to create channels containing the gold precursor *via* a self-assembly process, which upon photopolymerization created stable hexagonally closed path regions filled with gold nanoparticles. Glucose oxidase was then physisorbed on to the electrode and glucose oxidation was monitored by cyclic voltammetry (Figure 5A). An increase in anodic current was observed with a peak at 0.47 V (vs. Ag/AgCl), attributed to the oxidation of glucose. Song's glucose sensor utilized the powerful dispersing abilities of PILs to create stable graphene-PIL colloid in water. This suspension was then used to create a thin graphene film on a glassy-carbon electrode, which was subsequently treated with glucose oxidase (Figure 5B). A decrease in the reduction current for $O_2$ saturated solutions of glucose was detected and attributed to glucose consuming oxygen at the electrodes surface. With higher glucose concentrations (2.3 – 7.3 mM), a stronger decrease in the reduction current was observed. Alternatively, the $H_2O_2$ byproduct of glucose oxidation can be used to measure glucose levels, as demonstrated by Mecerreyes *et al.*.[85] Glucose oxidase was imbedded in PIL microgels (Figure 5C) and suspended in a glucose solution, and depending on the anion and glucose concentration, different currents were observed at 0.6 V (vs. SCE). More hydrophilic anions and greater glucose concentrations resulted in higher currents. These systems outperformed similar microgels composed of polyacrylamide and highlight the advantage of using PIL systems.

Apart from glucose, PIL-derived biosensors have served to detect dopamine, a ubiquitous neurotransmitter found among most multicellular organisms.[86] In humans this molecule plays a key role in the brain and central nervous system and underlies many fundamental biochemical pathways.[87] Thus its *in vivo* detection is important from a biomedical and research perspective. In a similar approach outlined previously, Song *et al.* used a graphene-oxide/polypyrrole/PIL composite as a highly sensitive dopamine sensor.[88] The PIL was polymerized off of the graphene-oxide/polypyrolle nanosheets and then cast on a glassy carbon electrode (Figure 4D).

The oxidation of dopamine at 0.4 V (vs. Ag/AgCl) was detected even in the presence of a 1500x excess of ascorbic acid (0.18 V), a case that is typically found in *in vivo* systems. Curiously, if they use graphene-oxide/polypyrrole sheets without PIL functionalization, the two oxidation events could not be distinguished. The presence of the PIL was paramount for the electrodes' function by effectively dispersing the nanosheets and allowing for good deposition on the electrode. This improved its sensitivity and demonstrates the secondary, yet significant role PILs play in such sensors.

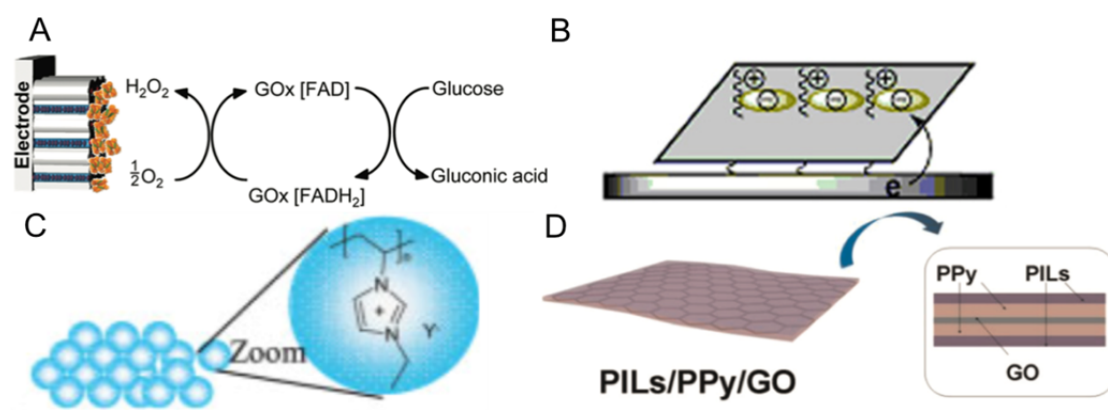

**Figure 5:** Various biosensor designs, including A) PIL-gold nanoparticle composites, B) PIL-graphene nanosheets, C) Glucose oxidase embedded PIL microgels for glucose sensing, and D) PIL/polypyrrole/grapheneoxide nanosheets for dopamine detection.

**Other Sensors**

The scope of PIL-derived sensors is under expansion and has been applied to various other system. Electrochemical sensors based on graphene/PIL composites have been developed for applications in hydrogen peroxide,[89] Sudan I,[90] and bisphenol A[91] sensing. In the first two cases, the PIL acts to stabilize and disperse the composites before deposition on the electrode. For hydrogen peroxide, they found that the introduction of the PIL resulted in much lower overpotentials during oxidation, allowing for easier detection. For Sudan I sensing, the presence of the PIL dramatically improved the oxidation current by promoting adsorption of the analyte on the PIL-functionalized electrode. This lowered the detection limits even further, to approximately $10^{-8}$ mol/L. These two examples demonstrate how the dispersing power of PILs can lead to better sensor design, and thus greater device performance. To avoid dispersions or deposition steps, the PIL may be synthesized directly on the electrode. Electrochemical polymerization has some benefits, such as the ability to form thin dense films and prevent the introduction of initiator species. A pyrrole-functionalized PIL was electropolymerized and used to detect bisphenol A.[91] They found that the PIL film possessed a porous structure which enhanced the sensitivity for analyte detection. Compared to an unfunctionalized, electrode, the PIL-coated sensor displayed 5x greater current for bisphenol A oxidation. This pronounced difference results from the accumulation of bisphenol A on the PIL, which in turn leads to a greater observed current. The authors found their sensor is comparable to other detection methods, such as HPLC, and reliable for practical applications. PILs have also been used as a temperature sensor through the exploitation of a fluorescent anion.[92] Self-assembled nanoparticles comprised of N-isopropylacrylamide and fluorescent IL moieties formed aggregates upon heating above the low critical solution temperature (LCST), resulting in enhanced fluorescence. Their approach yielded a highly sensitive reusable sensor for temperature in the LCST region.

**Conclusions**

The responsive nature of PILs towards a multitude of stimuli has been used in sensing applications and has provided quantitative and specific information for analyte detection. The significant investment in understanding the physical and chemical properties of ILs has transferred to PILs, and now to sensing applications. The tunable nature of the IL structure will allow for further development of PIL sensing and responsive capabilities. Moving from the lab to a real world application however is a big step that has yet to be taken in this field. While exploring the sensing capabilities of PILs is necessary, future work should also focus on the versatility and longevity of these sensors in field applications. Sensitivity and selectivity is only a prerequisite for a practical real-world sensor, as it must also withstand harsh operating conditions and have its limits/capabilities fully characterized for user operation. Entry of any new sensor into the commercial sector must compete with current technologies, which have been established as the benchmark. We believe the shear diversity PILs possess is its greatest advantage, and this manifests itself in the sensing technology discussed in this review. In the future niche applications for PIL sensors will emerge and their exploitation will then allow for the creation of a practical device.

**Acknowledgment**

R. G. and J. Y. thank the Max Planck society and the ERC (European Research council) Starting Grant with project number 639720—NAPOLI for the financial support. M. A. acknowledges financial support from the Marie Curie Actions of EU's 7$^{th}$ Framework Program under REA grant agreement no. 289347.

**Permissions**

**Figure 2A and 2B-** Reproduced from Ref 36 with permission of The Royal Society of Chemistry

**Figure 3A and 3B-** Reproduced from Ref 63 with permission of The Royal Society of Chemistry

**Figure 5A**- Reprinted with permission from S.Lee, B.S. Ringstrand, D. A. Stone, M. A. Firestone, ACS *Appl. Mater. Interfaces*, **2012,** *4*, 2311. Copyright 2012 American Chemical Society.

**Figure 5B-** Reprinted from Publication Q. Zhang, S. Wu, L. Zhang, J. Lu, F. Verproot, Y. Liu, Z. Xing, J. Li, X.M. Song, *Biosens. Bioelectron.* **2011**, *26*, 2632. with permission from Elsevier.